\documentclass[pdflatex,sn-mathphys-num,12pt]{sn-jnl}
\usepackage{graphicx}%
\usepackage{multirow}%
\usepackage{array}%
\usepackage{tabularx}%
\usepackage{amsmath,amssymb,amsfonts}%
\usepackage{bm}%
\usepackage{amsthm}%
\usepackage{mathrsfs}%
\usepackage[title]{appendix}%
\usepackage{xcolor}%
\usepackage{microtype}%
\usepackage{textcomp}%
\usepackage{manyfoot}%
\usepackage{booktabs}%
\usepackage{algorithm}%
\usepackage{algorithmicx}%
\usepackage{algpseudocode}%
\usepackage{listings}
\usepackage{cleveref}
\usepackage{docmute}
\usepackage{setspace}

\providecommand{\stopgrad}{\operatorname{sg}}

\newcommand{\R}{\mathbb{R}}

\begin{document}

\title{{S2S-JEPA: Predicting the Predictable at Subseasonal-to-Seasonal Timescales}}

\author*[1]{\fnm{Chenyu} \sur{Dong}}\email{chenyu.dong@u.nus.edu}

\author*[1]{\fnm{Gianmarco} \sur{Mengaldo}}\email{mpegim@nus.edu.sg}

\affil*[1]{\orgdiv{Department of Mechanical Engineering, College of Design and Engineering}, \orgname{National University of Singapore}, \orgaddress{\street{9 Engineering Drive 1}, \postcode{117575}, \country{Singapore}}}

\abstract{The subseasonal-to-seasonal (S2S) timescale, roughly from two weeks to two months ahead, is a critical forecast window for sectors such as agriculture, energy, and water management.
Yet, it is widely known as the `predictability desert'.
Recent AI weather models excel up to two weeks ahead but deteriorate beyond, largely because they are trained to predict fine-scale details that are neither predictable nor essential at S2S timescales.
We argue that a more physically grounded objective is to forecast only the slowly varying components that remain predictable.
Computer vision reached the same conclusion with the Joint-Embedding Predictive Architecture (JEPA), which predicts in latent space, discarding unpredictable details.
In this work, we introduce S2S-JEPA, which brings the JEPA paradigm to S2S forecasting.
It is tailored to this task through design elements from state-of-the-art AI weather models.
S2S-JEPA achieves comparable skill to the gold-standard ECMWF physics-based ensemble and surpasses it on multiple metrics at weeks 5 to 6.
}


\maketitle
\doublespacing

\section{Introduction}\label{intro}
Subseasonal-to-seasonal (S2S) prediction, which aims to forecast weather conditions from roughly two weeks to two months ahead, has long been a central challenge in atmospheric science.
The S2S timescale is a critical forecast window for decision-making in weather- and climate-sensitive sectors, such as agriculture, energy, and water management~\cite{white2017potential}.
It sits between medium-range weather forecasting (up to two weeks ahead) and seasonal prediction, and it is widely known as the `predictability desert'~\cite{brunet2010collaboration,mariotti2018progress}.
S2S prediction is particularly challenging because it extends beyond the timescale over which atmospheric initial conditions guide forecasts, yet remains too short for slower components, such as sea surface temperature, soil moisture, and large-scale climate modes, to fully dominate the forecast signal~\cite{merryfield2020current}.

Traditional physics-based numerical prediction systems, built on the numerical integration of the governing equations of atmospheric dynamics and their coupling with land, ocean, and sea-ice components, form the backbone of operational S2S prediction~\citep{vitart2017subseasonal}.
These forecasts are physically grounded, but integrating large coupled Earth-system models is computationally expensive.
This is particularly limiting at S2S lead times, where ensemble prediction is essential for isolating weak but potentially predictable signals from internal atmospheric variability.

Meanwhile, data-driven models have rapidly advanced medium-range weather forecasting, with many breakthroughs closely mirroring progress in artificial intelligence~\cite{camps2025artificial}.
This architectural transfer has been facilitated by the fact that most recent AI weather models are trained on gridded reanalysis data, which casts forecasting as a dense spatiotemporal prediction task that resembles computer-vision problems such as video prediction.
Neural operators in FourCastNet~\citep{pathak2022fourcastnet}, hierarchical Transformers in Pangu-Weather and FuXi~\citep{bi2023accurate,chen2023fuxi}, graph neural networks in GraphCast~\citep{lam2023learning} and diffusion models in GenCast~\citep{price2025probabilistic} have successively been adapted to global atmospheric prediction.
More recently, foundation models such as ClimaX~\citep{nguyen2023climax}, Aurora~\citep{bodnar2025foundation} and Prithvi WxC~\citep{schmude2024prithvi} have been pretrained on large-scale reanalysis and other meteorological datasets and then adapted to downstream variables, resolutions, and lead times.

Despite their strong performance at medium-range lead times, these AI weather models often struggle to extend their skill into the S2S range~\citep{lehmann2026can,nathaniel2024chaosbench}.
A key limitation lies in their training objective.
By requiring the model to reconstruct the full atmospheric state on the grid, these approaches devote capacity to details that become fundamentally unpredictable at longer lead times, which drives forecasts beyond two weeks toward smoothed fields with little skill over climatology.
Data-driven approaches to S2S prediction have been explored from several directions~\cite{weyn2021sub,cohen2019s2s,chen2024machine}.
One line of work treats the problem as statistical correction, learning to post-process dynamical ensemble outputs or to predict targeted quantities from selected climate drivers~\citep{cohen2019s2s,mouatadid2023adaptive}.
Other studies adapt the modeling paradigm of medium-range forecasting to subseasonal horizons, as in FuXi-S2S~\citep{chen2024machine} and TianQuan-S2S~\citep{li2026tianquan}, pretrain with masked reconstruction on reanalysis data, as in Baguan~\citep{niu2025utilizing}, or generate future state sequences in the latent space of a variational autoencoder, as in OmniCast~\citep{nguyen2025omnicast}.
In all of these models, however, the learning objective remains centered on reproducing grid-level details of the atmospheric state, either directly or through a latent space trained by reconstruction.
We argue that an objective better aligned with the physics of the Earth system is to forecast the slowly varying components that remain predictable at S2S timescales, such as tropical intraseasonal oscillations, stratospheric variability, and persistent land and oceanic anomalies, rather than fine-scale details that do not.

Remarkably, computer vision, the field that has repeatedly driven architectural innovation in AI for weather research, reached the same conclusion from a different starting point.
Masked prediction has become a dominant paradigm in self-supervised learning, with models reconstructing missing content from visible context~\citep{devlin2019bert,he2022masked}.
Joint-Embedding Predictive Architectures (JEPA) recast this idea in representation space, where the model infers the representations of hidden targets from context instead of reconstructing raw observations~\citep{lecun2022path,assran2023self}.
The formulation has since been applied to masked image regions~\citep{assran2023self}, spatiotemporal video blocks~\citep{bardes2024v,assran2025v}, and world models that forecast future latent states for planning and control~\citep{zhang2026hierarchical,maes2026leworldmodel,sun2026vla}.
Imagine watching a tree swaying in the wind~\citep{lecun2022path}: you neither track the motion of every leaf nor could you predict it; instead, what you perceive and anticipate is the higher-level structure of the scene—the tree continuing to sway with the wind. 
JEPA builds on this intuition. 
Rather than predicting the raw signal in all its unpredictable detail, it predicts a representation that captures what is predictable and sufficient for understanding what happens next.
This philosophy aligns well with the physical character of S2S predictability, where skill resides in a few slow, large-scale modes rather than in the full atmospheric state.

Motivated by this correspondence, we reformulate S2S forecasting as latent-space prediction, in which the model is tasked with forecasting the evolution of the Earth system's predictable structure.
In this work, we present S2S-JEPA, a fully data-driven S2S model built on the joint-embedding predictive principle.
Recent work has begun to explore JEPA for regional atmospheric prediction~\citep{anirudh2026tracing}. 
However, to our knowledge, S2S-JEPA is the first to extend this principle to global subseasonal forecasting, learning a predictive representation of the Earth system weeks into the future.
S2S-JEPA adapts the encoder-predictor architecture of V-JEPA~\citep{assran2025v,bardes2024v} from video to the Earth system, encoding daily snapshots of 79 atmospheric, land-surface and oceanic variables relevant to S2S predictability in place of frames of RGB pixels.
The training formulation, in contrast, is tailored to the physics of S2S prediction.
The model forecasts the future weekly latent states of the Earth system, so that S2S-timescale predictability itself supervises the encoder and the learned representation is forced to extract the slowly evolving signals that carry S2S predictability.
S2S-JEPA also adopts design choices that have proven effective at the medium range~\citep{nguyen2023climax,bodnar2025foundation,nguyen2024scaling,bi2023accurate}, as detailed in Section~\ref{sec:method}.
It delivers forecast skill competitive with the ECMWF operational S2S ensemble~\citep{vitart2017subseasonal}, the gold standard of subseasonal prediction, at weeks 5 to 6 and at a fraction of the computational cost.
The model trains and runs on a single GPU, whereas operational S2S forecasts require ensemble integrations of a physics-based model on dedicated supercomputing infrastructure.

\section{Method}
\label{sec:method}
\subsection{Problem setup}
\label{sec:setup}
We focus on the operational S2S forecasting task adapted from the ECMWF AI Weather Quest~\citep{loegel2025ai}.
Given an initialization day $t_0$, the task is to deliver probabilistic forecasts of weekly-mean 2-meter temperature (T2M), weekly accumulated precipitation (TP), and weekly-mean sea-level pressure (MSLP) for the following $K = 6$ weeks.
We refer to these three weekly quantities as the forecast products.
Temperature and precipitation are evaluated over land and sea-level pressure globally, in accordance with AI Weather Quest~\citep{loegel2025ai}.

Our models operate on $V = 79$ atmospheric, land-surface and oceanic variables from the ERA5 reanalysis~\citep{hersbach2020era5}, regridded to $1.5^\circ$ ($H \times W = 120 \times 240$; see Tab.~\ref{tab:variables} for the full list).
All variables are expressed as standardized anomalies with respect to a day-of-year climatology computed over 1979--2018 (see Appendix~\ref{app:data} for details).
This formulation isolates the forecast problem from the known seasonal cycle, focusing the model on the departures from climatology that constitute predictability at S2S timescales.

Let $X_t \in \R^{V \times H \times W}$ denote daily-mean anomaly fields of day $t$.
We group consecutive days into weeks of $T = 7$ days, indexed relative to the initialization day $t_0$.
Week $k$ is defined as
\begin{equation}
X^{(k)} = \left(X_{t_0+7k-6}, \dots, X_{t_0+7k}\right) \in \R^{T \times V \times H \times W}, \qquad k = 0, 1, \dots, K,
\label{eq:weeks}
\end{equation}
so that $X^{(0)}$ contains the seven days ending on $t_0$, and $X^{(k)}$ with $k \geq 1$ contains the $k$-th week after initialization.
The model input is the context week $X^{(0)}$, and $X^{(1)}, \dots, X^{(K)}$ are the weeks to be forecast.
The dependence of $X^{(k)}$ on $t_0$ is omitted to lighten the notation.
This input is longer than in prior S2S models, which condition on two to five daily states~\citep{chen2024machine,li2026tianquan}.
A full week of input allows slowly evolving signals, such as the Madden--Julian Oscillation, to be represented with their tendency as well as their instantaneous state.
It also has the same temporal extent as the target weeks, so that context and target weeks can be embedded by the same encoder.
S2S-JEPA does not predict the forecast products directly.
Its prediction target is the latent representation of each future week $X^{(k)}$, $k = 1, \dots, K$, which encodes all $V$ variables.
The three forecast products are obtained from these latent states by separate readout modules that do not influence the representation (Section \ref{sec:readout}).
We train on 1979–2018, validate on 2019–2021, and test on the 102 initializations from December 2023 to November 2025, a period for which the 101-member ECMWF extended-range ensemble is available.
The climatology defining the anomalies is computed on the training years alone, so no information from the evaluation period enters the problem definition, avoiding information leakage.
\begin{figure}[h!]
\centering
\includegraphics[width=\linewidth]{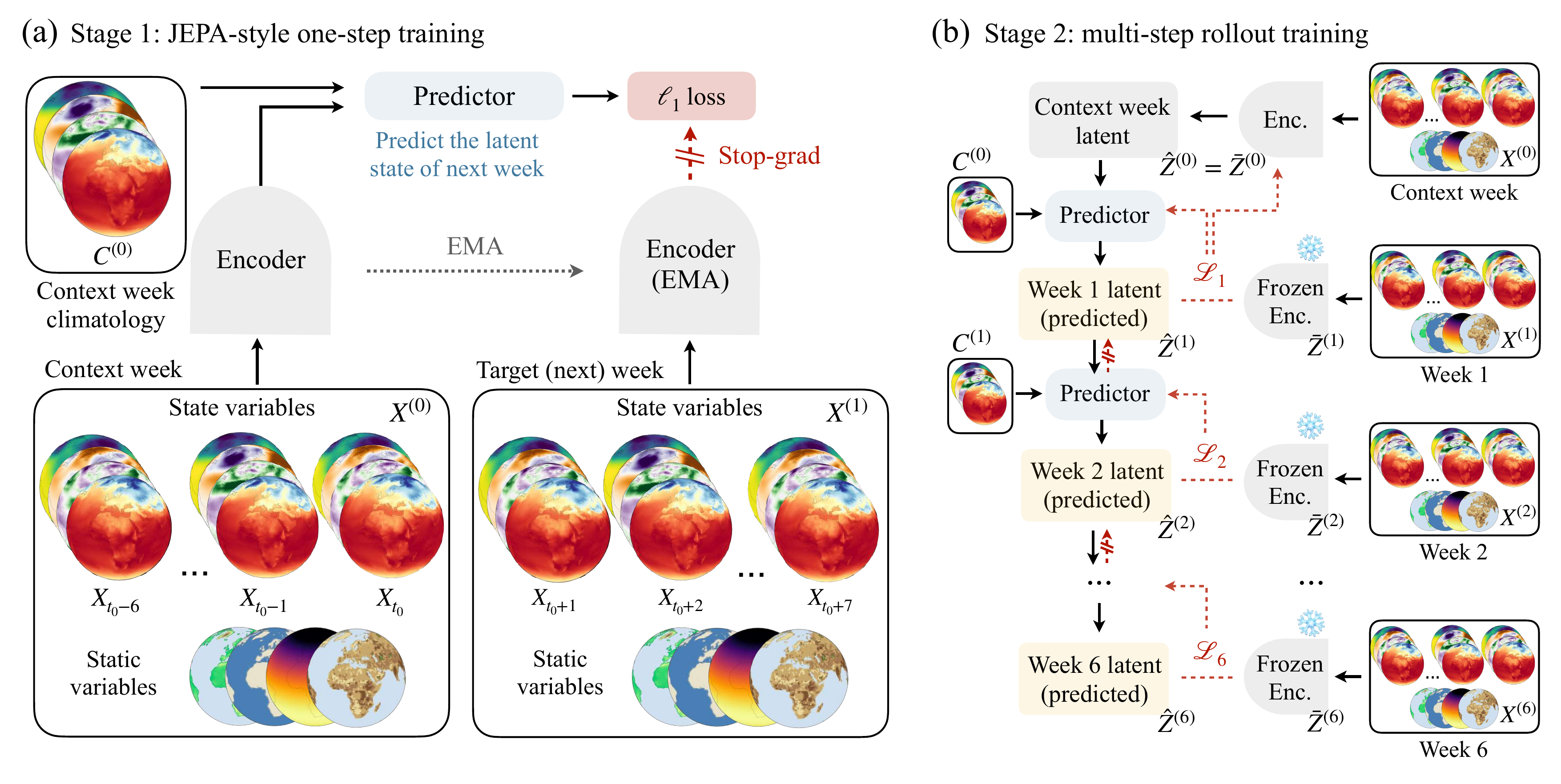}
\caption{Two-stage training of S2S-JEPA.
(a) Stage 1: one-step latent prediction.
The encoder embeds the context week $X^{(0)}$ (daily anomaly fields and static variables) into a latent state, and the predictor, conditioned on the context-week climatology $C^{(0)}$, predicts the latent state of the next week.
The prediction is matched under an $\ell_1$ loss to the embedding of the true next week under an EMA target encoder, which receives no gradient.
(b) Stage 2: multi-step rollout training.
Starting from the context-week latent, the predictor rolls out week-by-week predictions up to week 6, conditioned at each step on the climatology of the input week $C^{(k-1)}$, and each predicted latent is matched to the now-frozen target encoder's embedding of the observed week ($\mathcal{L}_1,\dots,\mathcal{L}_6$).
Gradients do not traverse the rollout chain; the context encoder is updated through the first step only.
}
\label{fig:schematic}
\end{figure}

\subsection{Joint-embedding predictive formulation and two-stage training}
\label{sec:jepa}

Joint-embedding predictive architectures~\citep{lecun2022path,assran2023self} learn by predicting the representation of a hidden part of a signal from that of a visible part.
A critical challenge in joint-embedding learning is representation collapse, where all inputs are mapped to the same or nearly identical embedding.
The target representations are therefore produced by a slowly evolving teacher, an exponential moving average of the online encoder, which stabilizes training and helps prevent this trivial solution.
S2S-JEPA treats the past week of the Earth system as the context and the unseen future week as the hidden target.
Training proceeds in two stages, the first on one-step prediction of the following week, the second on autoregressive rollouts across the full forecast range (Figure~\ref{fig:schematic}).
Separating the two stages allows the representation to be learned under a one-step objective before the predictor is trained on its own outputs.
The separation also proved necessary in practice, since training on rollouts from the start destabilized the representation and led to its rapid collapse in our early experiments.

\textbf{Stage 1: One-step latent prediction.}
In stage 1, an encoder $E_\theta$ maps the context week to a latent state $Z^{(0)} = E_\theta(X^{(0)}) \in \R^{N\times d}$ of $N$ tokens, and a predictor $P_\phi$ advances the latent state by exactly one week,
\begin{equation}
\hat{Z}^{(1)} = P_\phi\big(Z^{(0)},\, C^{(0)}\big),
\label{eq:onestep}
\end{equation}
where $C^{(0)}$ denotes the climatology of the context week.
The climatology conditions the predictor rather than the encoder, so that the learned representation cannot shortcut the prediction task by encoding a seasonal prior shared with the EMA target.
The regression target $\bar{Z}^{(1)} = E_{\bar\theta}(X^{(1)})$ is produced by a target encoder $E_{\bar\theta}$ applied to the true future week. 
The parameters of the target encoder are an exponential moving average of those of the online encoder, $\bar\theta \leftarrow m\,\bar\theta + (1-m)\,\theta$, where the momentum $m$ increases linearly from 0.999 to 1 over the course of Stage 1.
The encoder and the predictor are trained from scratch with the $\ell_1$ regression of $\hat{Z}^{(1)}$ to $\bar{Z}^{(1)}$, with tokens weighted by the cosine of latitude. 
The loss is computed entirely in latent space, and no reconstruction of physical fields contributes to the training of the encoder or the predictor.
Collapse of the representation is monitored during training through the deterministic readout head attached to the latent states (see Section~\ref{sec:readout} for details).

\textbf{Stage 2: Multi-step rollout training.}
Stage 1 leaves a distribution mismatch, as the predictor is trained on encoder outputs but rolled out on its own.
Stage 2 aims to close this gap by training the predictor on its own rollouts.
The multi-step objective does more than that, for as the rollout lengthens only signals that genuinely propagate through the latent dynamics can still be predicted, so the predictor devotes its capacity to the predictable components.
Concretely, starting from $\hat{Z}^{(0)} = Z^{(0)}$, the predictor is rolled out for $K=6$ steps, and each rolled-out state is matched to the target encoding of the true week at the same date,
\begin{equation}
\mathcal{L}_{Stage2} = \frac{1}{K}\sum_{k=1}^{K}
\ell\Big(P_\phi\big(\stopgrad\big(\hat{Z}^{(k-1)}\big),\, C^{(k-1)}\big),\; \bar{Z}^{(k)}\Big),
\label{eq:stage2}
\end{equation}
where $C^{(k-1)}$ denotes the climatology of the week being advanced from and $\stopgrad(\cdot)$ is the stop-gradient operator (Figure~\ref{fig:schematic}b).
The stop-gradient is omitted for $k=1$, the one step whose input is the encoder output itself, leaving an open path through which the encoder adapts the context representation to its role as the starting point of a rollout (Figure~\ref{fig:schematic}b).
Every predicted state is detached before it feeds the next step, so no gradient traverses the rollout chain.
The predictor thus receives $K$ one-step gradients evaluated on its own rollout distribution, and the target encoder is kept frozen so that the regression targets remain fixed throughout this stage.

Balancing one-step and multi-step objectives is known to be delicate for autoregressive weather models~\citep{nguyen2024scaling,bi2023accurate,chen2023fuxi,chen2024machine}.
We therefore sequence the two rather than balance them, borrowing the curriculum practice of medium-range models that train on short rollouts before long ones~\citep{lam2023learning,chen2023fuxi}.
Within Stage 2, gradients are not propagated through the rollout chain, so that late targets cannot reshape the early states from which they were produced.
Each step instead receives a one-step gradient evaluated on the rollout distribution, the `pushforward trick' that Aurora adopts for its roll-out fine-tuning~\citep{brandstetter2022message,bodnar2025foundation}.

\subsection{Model architecture}
\label{sec:arch}
The model components are drawn from the weather-modeling literature.
Their description is therefore kept brief, and the details are given in Appendix~\ref{app:model}.
In the encoder, a grouped weather embedding~\citep{nguyen2023climax,bodnar2025foundation} tokenizes the 79 variables into $N=4000$ tokens on five latent levels, from the ocean-land boundary through the near-surface state to the stratosphere, and a Swin Transformer backbone~\citep{liu2021swin,bi2023accurate} processes them with windows spanning all levels.
The predictor uses one learnable query for each output token, with global cross-attention to the encoded state to account for distant dependencies such as teleconnections, and windowed self-attention among the queries for local information exchange.
It outputs a full latent state in the same token space as the encoder, so that prediction and target can be compared directly under the joint-embedding loss.
Its only conditioning is seasonality, an input that prior S2S work has found effective~\citep{li2026tianquan}.
The climatology and top-of-atmosphere insolation of the input week modulate every token through FiLM layers~\citep{perez2018film}.

\subsection{Deterministic and probabilistic readout}
\label{sec:readout}
Since S2S-JEPA predicts latent states, the forecast products are obtained from them by two readout modules.
The deterministic readout provides a single estimate of the weekly anomalies, and the probabilistic readout generates an ensemble of forecasts, from which the quintile probabilities required by the task are computed.
In both cases, no gradient reaches the encoder or the predictor, so that the readouts do not influence the representation.

The deterministic readout is a lightweight convolutional decoder that maps each latent state to the weekly-mean anomalies of all 79 variables.
It is trained alongside the backbone on detached latent states, which also allows it to monitor training.
Its output serves as the deterministic forecast in Sec.~\ref{sec:exp} and as the basis for model selection.

The probabilistic readout follows the approach of ArchesWeatherGen~\citep{couairon2026archesweathergen}, which augments a deterministic forecasting model with a diffusion-based generative model trained to sample the residual distribution the deterministic forecast cannot express.
One head per forecast product is trained in the EDM framework~\citep{karras2022elucidating,price2025probabilistic} on the standardized residual, conditioned on the deterministic forecast, the predicted latent tokens through cross-attention, the lead time and the day of year (see Appendix~\ref{app:model}).
This diffusion-based generative model is trained separately, with the backbone of S2S-JEPA frozen.
At inference, 50 samples per target week form the ensemble, and quintile probabilities are the fractions of members falling into each category.

\subsection{Evaluation protocol}
\label{sec:eval}
We evaluate against the ECMWF extended-range ensemble from the S2S database~\citep{vitart2017subseasonal}, using its 101-member real-time forecasts and the 11-member reforecasts of the preceding 20 years.
The test set consists of the 102 Wednesday initializations between December 2023 and November 2025 for which both systems are available, and ERA5 weekly aggregates serve as ground truth for both.
Deterministic skill is measured by the temporal correlation coefficient (TCC) of anomalies at each grid point, computed over all test initializations.
We report TCC both as maps and as area-weighted averages, taken over land for temperature and precipitation and globally for sea-level pressure.
Probabilistic skill is measured by the fair ranked probability skill score (fair RPSS) of quintile forecasts, the metric of the ECMWF AI Weather Quest~\citep{loegel2025ai}.
Quintile boundaries are derived for each system from its own 20-year hindcast climate and for the observations from the ERA5 record over the same years, following FuXi-S2S~\citep{chen2024machine}; this removes model bias and drift from the comparison.
For 2-metre temperature, a linear trend fitted separately to the forecasts and to the observations is removed before the boundaries are computed, so that the warming trend does not count as skill.
Unlike FuXi-S2S, we do not detrend mean sea-level pressure and precipitation in the main text, for which the long-term trends are relatively weak.

\section{Results}\label{sec:exp}
\subsection{Deterministic skill}
Following the evaluation protocol of Sec.~\ref{sec:eval}, we compare S2S-JEPA against the operational ECMWF extended-range ensemble.
Tab.~\ref{tab:tcc} summarizes the deterministic skill (TCC) of both systems at weeks 3--6, with ECMWF ahead at week 3 and S2S-JEPA matching or exceeding the ensemble mean from week 4 onward for temperature and from week 5 for pressure.
\begin{table}[h!]
\centering
\caption{Deterministic skill at sub-seasonal lead times (weeks 3--6) on the 102 common Wednesday initializations (13 Dec 2023--19 Nov 2025): temporal correlation coefficient (TCC) of weekly-mean anomalies, area-weighted over land for T2M and TP and over the globe for MSLP.
ECMWF is the 101-member ensemble mean; S2S-JEPA is a single deterministic forecast. 
Anomalies are computed relative to the 2004--2023 ERA5 daily climatology.
Best values per column are shown in bold.
An upward arrow marks where S2S-JEPA is significantly better than ECMWF, a downward arrow where it is significantly worse, based on a bootstrap significance test over initializations (Appendix~\ref{app:significance}).}
\label{tab:tcc}
\small
\begin{tabularx}{\textwidth}{ll*{4}{>{\centering\arraybackslash}X}}
\toprule
 & & \multicolumn{4}{c}{Lead (week)} \\
\cmidrule(lr){3-6}
Variable & System & 3 & 4 & 5 & 6 \\
\midrule
\multirow{2}{*}{T2M} & ECMWF (ens.\ mean) & \textbf{0.349} & 0.244 & 0.202 & 0.181 \\
 & S2S-JEPA (ours) & 0.289$^{\downarrow}$ & \textbf{0.251} & \textbf{0.238}$^{\uparrow}$ & \textbf{0.230}$^{\uparrow}$ \\
\midrule
\multirow{2}{*}{MSLP} & ECMWF (ens.\ mean) & \textbf{0.467} & \textbf{0.338} & \textbf{0.288} & 0.260 \\
 & S2S-JEPA (ours) & 0.335$^{\downarrow}$ & 0.304$^{\downarrow}$ & 0.283 & \textbf{0.277} \\
\midrule
\multirow{2}{*}{TP} & ECMWF (ens.\ mean) & \textbf{0.270} & \textbf{0.204} & \textbf{0.168} & \textbf{0.161} \\
 & S2S-JEPA (ours) & 0.191$^{\downarrow}$ & 0.168 & 0.157 & 0.150 \\
\bottomrule
\end{tabularx}
\end{table}
\begin{figure}[h!]
\centering
\includegraphics[width=1\linewidth]{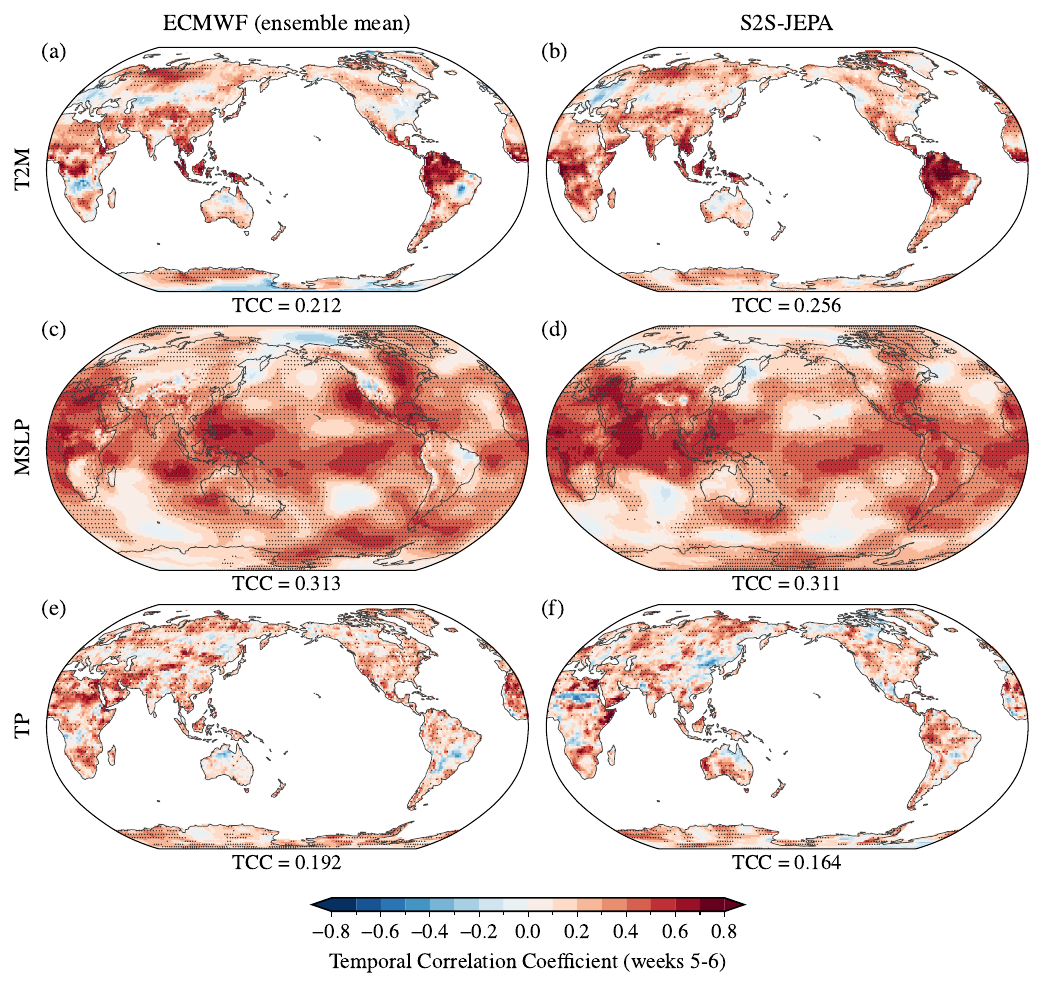}
\caption{Spatial distribution of deterministic skill at weeks 5--6.
Grid-point TCC of weekly-mean anomalies for the ECMWF ensemble mean (left) and S2S-JEPA (right), computed jointly over weeks 5 and 6 on the same initializations as Tab.~\ref{tab:tcc}.
Numbers below each panel give the area-weighted average, over land for T2M and TP and over the globe for MSLP.
Stippling marks grid points where the correlation is significant.} 
\label{fig:tcc}
\end{figure}
ECMWF stays ahead on precipitation, intrinsically the hardest of the three targets, across weeks 3--6, though the gap narrows as lead time increases.
Together, these results show a clear lead-time dependence, where S2S-JEPA becomes increasingly competitive as lead time grows, and by weeks 5--6, it matches the operational ensemble across all three variables, while achieving significantly better performance for 2-meter temperature.

To see where this skill resides, we further examine the spatial distribution of TCC at weeks 5--6, computed at each grid point as in Tab.~\ref{tab:tcc} using the two lead weeks jointly (Fig.~\ref{fig:tcc}).
For 2-meter temperature, the two systems largely agree on the spatial distribution of predictability, but S2S-JEPA shows higher skill over South America, Africa, India, and the Indochina Peninsula.
For sea-level pressure, the two area-averaged values are nearly identical, and the spatial patterns largely agree, with each system having regions where it performs better.
Precipitation skill is more scattered for both models, with ECMWF retaining an overall advantage and exhibiting significant skill over a broader area.

\subsection{Probabilistic skill}
We next evaluate probabilistic skill using the fair ranked probability skill score (RPSS) for quintile forecasts.
Tab.~\ref{tab:rpss} compares the 50-member S2S-JEPA EDM ensemble against the operational ECMWF ensemble at weeks 3--6.
As in the deterministic evaluation, ECMWF has a clear advantage at week 3 across all three variables, while S2S-JEPA becomes increasingly competitive at longer lead times.

For 2-meter temperature, S2S-JEPA matches or exceeds ECMWF from week 4 onward and is significantly better at week 6.
Notably, precipitation, for which S2S-JEPA has lower deterministic skill than ECMWF throughout weeks 3--6, shows strong probabilistic performance at longer lead times, matching ECMWF at week 4 and significantly outperforming it at weeks 5 and 6.
Sea-level pressure shows a different behavior, with ECMWF retaining higher fair RPSS throughout weeks 3--6, although the difference is no longer significant at the two longest lead times.
Overall, these results mirror the lead-time dependence observed for deterministic skill, while S2S-JEPA shows particularly strong long-range probabilistic performance for temperature and precipitation.

\begin{figure}[h!]
\centering
\includegraphics[width=1\linewidth]{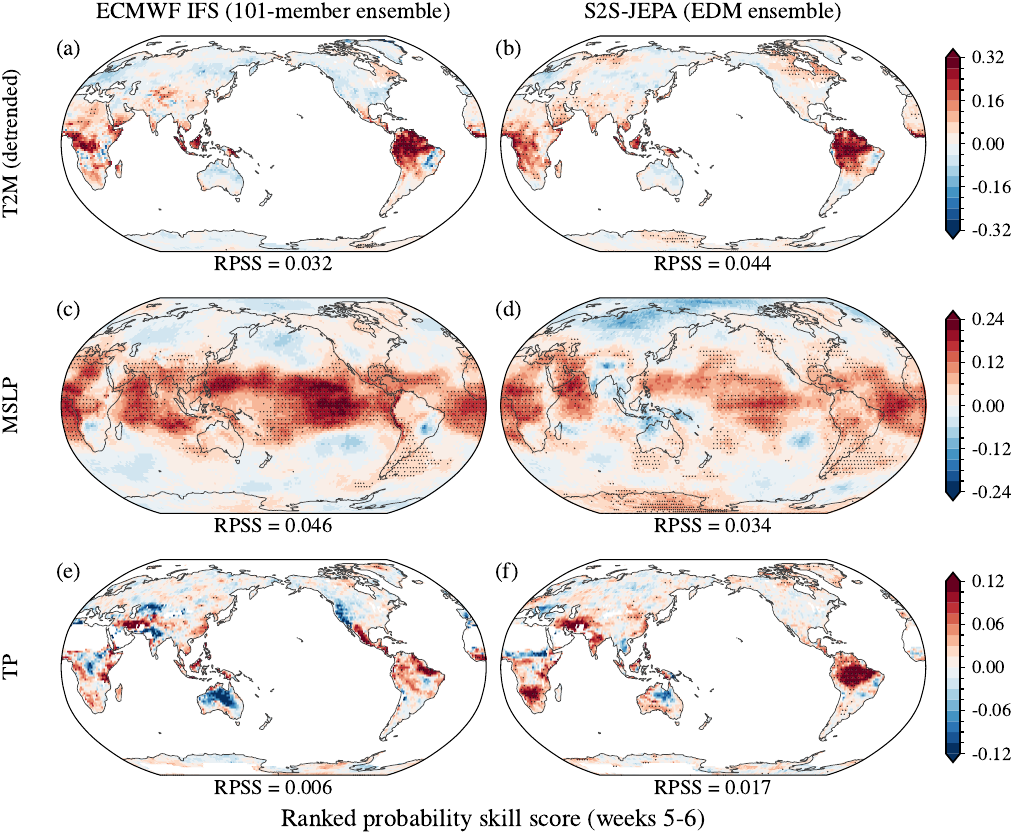}
\caption{As in Fig.~\ref{fig:tcc}, but for the RPSS of quintile forecasts relative to climatology.
For TP, grid points where the climatological quintiles are degenerate due to near-zero precipitation are masked.
}
\label{fig:rpss}
\end{figure}

The spatial distribution at weeks 5--6 further highlights these differences (Fig.~\ref{fig:rpss}).
For 2-meter temperature, both systems show similar spatial patterns of predictability, with strong skill over tropical South America, Africa, and the Maritime Continent.
\begin{table}[h!]
\centering
\caption{Probabilistic skill at weeks 3--6: fair RPSS of quintile probabilities relative to the uniform forecast under the FuXi-S2S protocol, with each system's quintile edges taken from its own 2004--2023 reforecasts (11 members per initialization). S2S-JEPA + EDM has 50 members. T2M is scored after removing each system's own linear trend and the observed trend; MSLP and TP are scored without detrending. TP grid points with a degenerate quintile range are masked.}
\label{tab:rpss}
\small
\begin{tabularx}{\textwidth}{ll*{4}{>{\centering\arraybackslash}X}}
\toprule
 & & \multicolumn{4}{c}{Lead (week)} \\
\cmidrule(lr){3-6}
Variable & System & 3 & 4 & 5 & 6 \\
\midrule
\multirow{2}{*}{T2M (detrended)} & ECMWF & \textbf{0.092} & 0.049 & 0.040 & 0.031 \\
 & S2S-JEPA + EDM & 0.079$^{\downarrow}$ & \textbf{0.058} & \textbf{0.049} & \textbf{0.043}$^{\uparrow}$ \\
\midrule
\multirow{2}{*}{MSLP} & ECMWF & \textbf{0.152} & \textbf{0.075} & \textbf{0.053} & \textbf{0.041} \\
 & S2S-JEPA + EDM & 0.078$^{\downarrow}$ & 0.052$^{\downarrow}$ & 0.044 & 0.035 \\
\midrule
\multirow{2}{*}{TP} & ECMWF & \textbf{0.048} & 0.021 & 0.008 & 0.005 \\
 & S2S-JEPA + EDM & 0.030$^{\downarrow}$ & 0.021 & \textbf{0.018}$^{\uparrow}$ & \textbf{0.015}$^{\uparrow}$ \\
\bottomrule
\end{tabularx}
\end{table}
S2S-JEPA additionally shows significant skill over regions including eastern Canada and the Arabian Peninsula, contributing to a higher area-averaged RPSS than ECMWF.
For sea-level pressure, ECMWF retains the higher global-average skill, with stronger positive RPSS over broad regions of the tropical and subtropical oceans.
The largest relative improvement is found for precipitation. 
Despite substantial spatial heterogeneity in both systems, S2S-JEPA achieves an area-averaged RPSS of 0.017, compared with 0.006 for ECMWF, with coherent regions of improvement across the globe.
We note that detrending is applied only to T2M (see Sec.~\ref{sec:eval}), as predicted warming coinciding with the pronounced long-term trend could otherwise be misinterpreted as genuine forecast skill over climatology.
For completeness, Appendix~\ref{app:trend} reports MSLP and TP with the same detrending procedure.
Detrending favors S2S-JEPA, which becomes comparable to ECMWF for MSLP and substantially better for TP at weeks 5--6, although the latter should be interpreted cautiously given the noisy trend estimates from the reforecasts.

\subsection{Sources of forecast skill}
We use controlled ablations to motivate the key design choices of S2S-JEPA, with all variants evaluated on the same test set under the protocol of Sec.~\ref{sec:eval} using TCC averaged across the three variables (Table~\ref{tab:ablation}).
In particular, we examine the forecasting formulation, the role of latent prediction, and the training strategy for autoregressive rollouts.

We first examine the choice of forecasting formulation for our S2S setting, as different formulations have been explored for S2S prediction in AI weather models, including autoregressive forecasting~\cite{chen2024machine} and direct prediction at prescribed lead times~\cite{li2026tianquan}.
In fact, we initially expected lead-conditioned forecasting to be better suited to the S2S timescale, given the potential accumulation of errors over repeated autoregressive steps and the strong performance demonstrated by TianQuan-S2S~\cite{li2026tianquan}.
For a controlled comparison, we consider two direct forecasting variants with the same backbone: a lead-conditioned model, with the target lead time (weeks 1--6) encoded and injected into each Transformer block through adaptive LayerNorm conditioning following Stormer~\cite{nguyen2024scaling}, and per-lead experts trained separately for each target week~\cite{li2026tianquan}.
Both variants use the same JEPA-style training objective, with the loss applied in latent space, isolating the forecasting formulation as the main difference.
\begin{table}[h]
\centering
\caption{
Ablations of the key design choices in S2S-JEPA.
Values report the mean TCC across T2M, MSLP, and TP at each lead time.
}
\label{tab:ablation}
\small
\setlength{\tabcolsep}{5pt}
\begin{tabularx}{\textwidth}{l*{6}{>{\centering\arraybackslash}X}}
\toprule
& \multicolumn{6}{c}{Lead (week)} \\
\cmidrule(lr){2-7}
Model & 1 & 2 & 3 & 4 & 5 & 6 \\
\midrule

\textbf{S2S-JEPA (final)}
& 0.609 & \textbf{0.335} & \textbf{0.271} & \textbf{0.241} & \textbf{0.226} & \textbf{0.219} \\

\midrule
\multicolumn{7}{l}{\textit{Forecasting formulation}} \\
\addlinespace[2pt]
Lead-conditioned
& 0.422 & 0.265 & 0.209 & 0.175 & 0.152 & 0.141 \\
Per-lead experts
& 0.620 & 0.299 & 0.232 & 0.208 & 0.198 & 0.187 \\

\midrule
\multicolumn{7}{l}{\textit{Prediction objective}} \\
\addlinespace[2pt]
S2S-JEPA (Stage 1 only)
& 0.620 & 0.295 & 0.205 & 0.176 & 0.160 & 0.155 \\
Supervised (Stage 1 only)
& \textbf{0.647} & 0.207 & 0.121 & 0.119 & 0.116 & 0.116 \\
Supervised (Stages 1 \& 2)
& 0.623 & 0.318 & 0.194 & 0.159 & 0.134 & 0.120 \\

\midrule
\multicolumn{7}{l}{\textit{Rollout training}} \\
\addlinespace[2pt]
Curriculum
& 0.598 & 0.313 & 0.260 & 0.222 & 0.210 & 0.207 \\
Mixed-horizon rollout
& 0.586 & 0.296 & 0.251 & 0.224 & 0.218 & 0.215 \\

\bottomrule
\end{tabularx}
\end{table}
The results nevertheless favor autoregressive forecasting.
The lead-conditioned model performs substantially worse across all lead times, while the per-lead experts achieve slightly higher skill at week 1 (0.620 versus 0.609) but fall behind from week 2 onward.
We attribute these differences in part to the opposite trade-offs during pretraining: lead conditioning requires a single model to accommodate prediction tasks across widely different timescales, whereas per-lead experts restrict each model to samples from a single target lead time, reducing the diversity of its pretraining data.
Autoregressive forecasting balances these two aspects by defining a consistent prediction task while retaining the diversity of training samples across lead times.

The autoregressive formulation brings us to the central hypothesis behind S2S-JEPA, namely that predicting in latent space, rather than directly regressing the forecast fields, is key to maintaining skill at the S2S timescale.
To test this directly, we train a supervised counterpart with the same encoder–predictor backbone, data, and training budget, replacing latent prediction with direct regression of the forecast fields.
The supervised counterpart predicts the seven daily anomaly fields of the following week, so that each predicted week has the same form as the model input and is fed back directly as the input of the next rollout step.
We first compare the two objectives after one-step training only (Stage 1 in Fig.~\ref{fig:schematic}), before either model is exposed to rollouts during training.
Interestingly, the supervised model achieves higher skill at week 1 than S2S-JEPA (0.647 versus 0.620), showing that direct supervised regression performs better when the task is limited to a single-week prediction.
This suggests that, while learning in latent space does not improve one-step performance, it yields representations with substantially better properties for autoregressive rollout.
The same pattern remains after rollout training (Stage 2).
Both models improve from week 2 to week 6 at the cost of a slight decrease at week 1, and S2S-JEPA retains substantially better skill from week 2 onward, although the supervised model remains better at week 1.
This indicates that rollout fine-tuning benefits both settings, but cannot compensate for the loss of rollout skill under direct supervised regression.
We interpret this gap in rollout skill as evidence for the \textit{predict the predictable} principle underlying S2S-JEPA.
With direct supervised learning, the pointwise loss requires the model to account for the full target state, including variability that becomes increasingly unpredictable at longer timescales. 
In contrast, JEPA shifts the prediction target to latent space, encouraging representations that emphasize the more predictable components of the Earth system.
We believe this is what allows skill learned from one-step prediction to persist naturally toward longer autoregressive leads.

We emphasize that this does not imply that direct field reconstruction is unsuitable for S2S prediction, as previous studies have demonstrated its effectiveness in S2S forecasting~\cite{chen2024machine,li2026tianquan}.
Rather, the week-to-week setup considered here is particularly well aligned with the spirit of S2S-JEPA, as the long input and target windows emphasize slowly evolving and more predictable modes, while reconstructing the full target fields over the same interval makes direct supervised learning substantially more challenging.

Finally, we examine the training strategy for the autoregressive model, comparing the rollout training strategies explored during model development.
Here, we report representative comparisons among two-stage training, curriculum learning, and mixed-horizon rollout training.
Two-stage training~\cite{pathak2022fourcastnet,bodnar2025foundation}, which first learns one-step prediction and then fine-tunes on rollouts, performs best at the S2S leads and is therefore adopted in the final S2S-JEPA (Stages 1--2 in Fig.~\ref{fig:schematic}).
Curriculum learning~\cite{chen2024machine,lam2023learning}, which progressively increases the rollout depth, shifts skill toward the earlier weeks at the expense of longer leads. 
Mixed-horizon rollout training~\cite{nguyen2024scaling}, which back-propagates through the rollout and jointly optimizes multiple forecast steps, performs slightly worse overall, although the gap becomes small at weeks 5--6.
Overall, these results favor explicit one-step pretraining followed by rollout fine-tuning for optimizing skill across the S2S range, leading us to adopt this strategy in S2S-JEPA.

\section{Discussion}\label{sec:discussion}
In this work, we introduced S2S-JEPA, a latent-space forecasting framework for S2S prediction.
The framework is motivated by a fundamental distinction between S2S and medium-range weather forecasting. 
As lead time increases, the precise evolution of individual weather systems becomes increasingly unpredictable, while slower and larger-scale components of the Earth system can retain useful predictability.
S2S-JEPA therefore shifts the learning objective from reproducing every detail of the future atmospheric state toward predicting what remains predictable, the same principle that JEPA formalizes for world models by placing the prediction objective in a learned representation space~\cite{lecun2022path}.
We hypothesized that it could translate into improved forecast skill. 
This hypothesis follows a broader lesson from data-driven weather forecasting, where the successful adaptation of ideas from computer vision has often relied on integrating architectural innovations with the underlying physics as well as on transferring them~\cite{pathak2022fourcastnet,bi2023accurate,lam2023learning,bodnar2025foundation,wang2026condensnet}.

In our formulation, S2S-JEPA first learns to predict the latent representation of the following week from the preceding week, and is then trained on autoregressive rollouts to extend prediction to six weeks.
Probabilistic ensembles are generated by coupling the latent predictor with a diffusion-based generative model.
We evaluate both deterministic and probabilistic skill against the operational ECMWF extended-range ensemble.
Across both settings, S2S-JEPA becomes increasingly competitive with ECMWF as lead time increases.
For deterministic forecasts, S2S-JEPA is significantly worse than the ECMWF ensemble mean at week 3, but by weeks 5 and 6 it matches it for MSLP and TP and achieves significantly higher TCC for T2M.
A similar transition emerges in probabilistic forecasting. Despite a clear disadvantage across all three variables at week 3, S2S-JEPA matches or exceeds ECMWF at longer lead times for T2M and TP, with significant improvements for T2M at week 6 and TP at weeks 5 and 6, while ECMWF retains an advantage for MSLP.
This lead-time dependence is further supported by the ablation against supervised learning, where the relative benefit of latent prediction likewise becomes more pronounced at longer rollout horizons.
Together, these observations are consistent with our hypothesis. 
We expect the benefits of latent prediction to emerge only as the detailed evolution of the weather loses predictability while large-scale structure persists.



Despite its competitive performance at longer lead times, S2S-JEPA remains substantially less skillful than ECMWF at weeks 1--3, constituting an important limitation of the current framework.
It is therefore also expected to be less competitive at shorter lead times than daily-resolution models such as FuXi-S2S~\cite{chen2024machine}, whose short-lead skill is comparable to the ECMWF ensemble.
This short-lead deficit may reflect a trade-off inherent in our formulation, where by emphasizing predictable information in latent space, together with the weekly temporal aggregation of both inputs and targets, the model may discard fine-grained information that remains valuable when the evolution of individual weather systems is still relatively predictable.
Understanding whether this trade-off is intrinsic to our latent prediction formulation, and whether short-lead skill can be improved through alternative task formulations without sacrificing longer-lead performance, is an important direction for future work.
We also acknowledge that S2S-JEPA does not outperform ECMWF across all variables and metrics even at longer lead times.
One possible explanation is that the latent prediction objective selectively preserves predictable information, but the information retained in the learned representation may not align equally well with every variable.
This issue may be less pronounced in video, where the input channels typically represent RGB values and are jointly interpreted as visual content.
At S2S timescales, by contrast, different physical variables can have substantially different levels and sources of predictability, potentially biasing the learned representation toward a certain subset of variables.

More broadly, as the first exploration of JEPA for S2S prediction, S2S-JEPA demonstrates the feasibility of this direction while leaving substantial room for further development.
One natural extension is to scale pretraining to larger and more diverse Earth-system datasets, following the scaling trend already established in Earth-system foundation models~\cite{nguyen2023climax,bodnar2025foundation}.
Another promising direction is to incorporate masking strategies commonly used in JEPA pretraining.
In particular, decades of S2S research have provided substantial prior knowledge about sources of subseasonal predictability, opening the possibility of \emph{physics-aware masking} that selectively exposes or masks known sources of predictability.
Such a strategy could encourage the model to learn physically meaningful predictive relationships while preventing reliance on simple shortcuts.
Meanwhile, explainability is increasingly important in data-driven weather and climate prediction, not only for building trust in forecasts but also for advancing scientific understanding~\cite{macmillan2025towards,wei2025xai4extremes}. 
Interpreting the learned representations with emerging XAI tools could reveal the physical drivers behind S2S-JEPA's predictions and offer new insights into the intrinsic predictability of the Earth system~\cite{dong2025time}.
Finally, rapidly evolving JEPA formulations, such as LeWorldModel~\citep{maes2026leworldmodel}, together with growing applications of JEPA across scientific domains, offer alternative approaches to learning latent dynamics and could provide useful building blocks for future S2S-JEPA models.
Together, these directions highlight representation-space prediction as a promising new avenue for extended-range Earth-system forecasting~\cite{camps2026bridging}.

\clearpage
\bibliography{references}  

\clearpage
\appendix

\section*{Appendix}

\section{Data}\label{app:data}

\subsection{Variable List}

For S2S-JEPA, we use daily-mean ERA5 fields on a global $1.5^\circ$ grid, with 79 time-varying input fields in total (Tab.~\ref{tab:variables}).
Beyond the atmospheric and surface variables commonly used in medium-range weather forecasting, our input state includes additional components relevant to subseasonal predictability, including soil temperature and moisture, ocean surface conditions, and an extended vertical coverage into the stratosphere up to 10\,hPa.
These additions are intended to better represent the slowly varying components and boundary conditions that can provide sources of predictability at S2S timescales.

\begin{table}[h]
\label{tab:variables}
\centering
\caption{Variables used by S2S-JEPA.
All fields are daily means of ERA5 on the global $1.5^\circ$ grid, and short names follow the ERA5 convention.
Pressure-level variables are taken at 1000, 925, 850, 700, 600, 500, 400, 300, 200, 100, 50 and 10\,hPa.
The column Fields gives the number of two-dimensional fields that a variable contributes, which adds up to 79 for the time-varying inputs.
The static fields enter the encoder and the insolation enters the predictor.
The forecast targets are t2m, msl and tp.}
\label{tab:variables}
\centering
\small
\begin{tabularx}{\textwidth}{@{}lXlr@{}}
\toprule
{\bf Type} & {\bf Variable} & {\bf Short name} & {\bf Fields} \\
\midrule
Pressure level & Geopotential                    & z              & 12 \\
               & Temperature                     & t              & 12 \\
               & U component of wind             & u              & 12 \\
               & V component of wind             & v              & 12 \\
               & Specific humidity               & q              & 12 \\
\addlinespace
Single level   & 2\,m temperature                & t2m            & 1  \\
               & 2\,m dewpoint temperature       & d2m            & 1  \\
               & Mean sea level pressure         & msl            & 1  \\
               & 10\,m U component of wind       & u10            & 1  \\
               & 10\,m V component of wind       & v10            & 1  \\
               & Total column water vapour       & tcwv           & 1  \\
               & Total precipitation             & tp             & 1  \\
               & Top net thermal radiation       & ttr            & 1  \\
\addlinespace
Land           & Soil temperature, 4 layers      & stl1 to stl4   & 4  \\
               & Volumetric soil water, 4 layers & swvl1 to swvl4 & 4  \\
               & Snow depth                      & sd             & 1  \\
\addlinespace
Ocean          & Sea surface temperature         & sst            & 1  \\
               & Sea ice area fraction           & siconc         & 1  \\
\midrule
\multicolumn{3}{@{}l}{Time-varying inputs in total} & 79 \\
\midrule
Static         & Orography                       & orog           & 1  \\
               & Land-sea mask                   & lsm            & 1  \\
               & Soil type                       & slt            & 1  \\
\bottomrule
\end{tabularx}
\end{table}

\subsection{Data Preprocessing}

All inputs are ERA5 \citep{hersbach2020era5} daily fields for 1979--2025 on a $1.5^\circ$ grid conservatively regridded from $0.25^\circ$; accumulated quantities (precipitation and top net thermal radiation) as daily totals, all others as daily means.
To account for the heavy-tailed distribution of precipitation, daily precipitation is transformed as $\log(1+x/\varepsilon)$, where $\varepsilon$ is the 10th percentile of positive daily totals in the training years ($1.3\times10^{-5}$\,m).
S2S-JEPA operates in anomaly space, with anomalies defined relative to a fixed day-of-year climatology computed exclusively from the training years 1979--2018. 
Note that this training climatology is distinct from the 2004–2023 verification climatology used to compute TCC.
For each variable and grid point, the climatological mean and standard deviation are estimated by pooling samples at day offsets ${-4,-2,0,2,4}$ around each calendar day and smoothing the resulting day-of-year statistics using the first four annual harmonics.
Anomalies are then computed as $(x-\mu_{\mathrm{doy}})/\sigma_{\mathrm{doy}}$ and clipped at $\pm10$ as a safeguard against extreme outliers.
The $-90^\circ$ row (zero area weight) is dropped, giving a $120\times240$ grid on which all forecasts and verification are carried out.

\section{Model Architecture Detail}\label{app:model}

\subsection{S2S-JEPA}

\paragraph{Encoder.}

The encoder maps the seven context days to a latent state through grouped embedding followed by a stack of Swin Transformer blocks.
Inspired by the variable-wise embedding of ClimaX~\citep{nguyen2023climax} and the latent-level aggregation of Aurora~\citep{bodnar2025foundation}, our grouped embedding partitions the 79 variables into 16 physically motivated groups, which are further organized into five latent levels (Tab.~\ref{tab:groups}).
For each group, the seven context days are stacked along the channel dimension, following the treatment of consecutive states in GraphCast and FuXi~\citep{lam2023learning,chen2023fuxi}, and embedded by a group-specific convolution with $6\times6$ patches.
Ocean and land variables form a boundary level; near-surface state and column variables form a surface level; and pressure-level variables are divided into lower-tropospheric (1000--700\,hPa), free-tropospheric (600--200\,hPa), and stratospheric (100--10\,hPa) levels, reflecting the distinct physical regimes and the broad physical structure of the Earth system.
Within each latent level, the resulting group embeddings are concatenated and projected to a single token per patch, to which sinusoidal level and positional embeddings are added.
This yields $N=5\times20\times40=4000$ tokens of width $d=512$.
The tokens are processed by six Swin Transformer blocks~\citep{liu2021swin} with three-dimensional windows in the spirit of Aurora~\citep{bodnar2025foundation}.
Each window spans all five levels and $4\times 4$ patches, so that vertical coupling is resolved in every layer.

\begin{table}[t]
\centering
\caption{The 79 ERA5 variables are first partitioned into 16 physically motivated groups, which are further organized into five latent levels.
Static inputs are appended to the groups indicated in the fourth column.
\emph{Total} denotes the number of distinct fields in each row, including dynamic variables and static inputs.
For each group, the seven input days are stacked channel-wise and embedded with a $6\times6$ patch convolution.
The resulting group embeddings within each latent level are then concatenated and projected to a single token per patch.}
\label{tab:groups}
\small
\begin{tabularx}{\textwidth}{XXXXr}
\toprule
Latent level & Group & Variables & Static Input & Total \\
\midrule
\multirow{2}{*}{Boundary} & Ocean & \texttt{sst}, \texttt{siconc} & Orography, land--sea mask, soil type & 5 \\
 & Land & \texttt{sd}, \texttt{stl1-4}, \texttt{swvl1-4} & Orography, land--sea mask, soil type & 12 \\
\midrule
\multirow{2}{*}{Surface} & Near-surface & \texttt{t2m}, \texttt{d2m}, \texttt{msl}, \texttt{u10}, \texttt{v10} & Orography, land--sea mask, soil type & 8 \\
 & Column & \texttt{tcwv}, \texttt{tp}, \texttt{ttr} & Orography, land--sea mask, soil type & 6 \\
\midrule
Lower troposphere & 1000, 925, 850, 700\,hPa & \texttt{q}, \texttt{t}, \texttt{u}, \texttt{v}, \texttt{z} at each level & Orography & 21 \\
Free troposphere & 600, 500, 400, 300, 200\,hPa & \texttt{q}, \texttt{t}, \texttt{u}, \texttt{v}, \texttt{z} at each level & -- & 25 \\
Stratosphere & 100, 50, 10\,hPa & \texttt{q}, \texttt{t}, \texttt{u}, \texttt{v}, \texttt{z} at each level & -- & 15 \\
\bottomrule
\end{tabularx}
\end{table}

\paragraph{Predictor.}

The predictor maps the encoded state to the latent representation of the following week.
It first applies two global self-attention blocks to globally contextualize the encoded state, which then serves as the memory for prediction. 
To construct the next latent state, one learned query is assigned to each of the $N=4000$ positions on the $5\times20\times40$ latent grid.
These queries are processed by eight predictor blocks, each consisting of global cross-attention to the encoded memory, windowed self-attention among the queries, and an MLP.
Global cross-attention allows each predicted token to access the entire current state, capturing long-range dependencies such as teleconnections, while windowed self-attention enables local information exchange among tokens of the predicted state.
Unlike predictors conditioned on lead time~\citep{nguyen2024scaling}, ours always advances the state by one week and is instead conditioned on the input week climatology.
The climatological mean and log-standard deviation of all variables at the input week, together with top-of-atmosphere insolation, are embedded following the same level structure as the encoder and modulate the three branches of every query block through FiLM layers~\citep{perez2018film}.
The predictor thus defines a non-autonomous operator with explicit seasonal forcing.
It predicts all five latent levels, including the boundary level, yielding a complete latent state that can be fed directly into the next prediction step.
During training, each step is trained against the teacher representation of its corresponding target week.

\subsection{Deterministic readout head}

A lightweight convolutional head decodes the latent state into physical fields and is shared across all rollout steps.
At each spatial patch, the tokens from the five latent levels are concatenated and passed through a $1\times1$ convolution, a $3\times3$ convolution, and a final $1\times1$ convolution with GELU activations.
The output channels are then rearranged into $6\times6$ pixels by pixel shuffle, recovering the $120\times240$ grid.
The head predicts standardized weekly-mean anomalies for all 79 variables, rather than only the three forecast targets, providing a diagnostic of how much information about the full Earth-system state is retained in the latent representation.
Precipitation is instead predicted as the weekly total in the logarithmic space, since averaging daily values after the log transform does not preserve the weekly total.

The readout head is trained jointly with the backbone but acts only as a probe: its latitude-weighted mean-squared-error loss is computed against the ground truth, and gradients are stopped at the latent state so that the readout loss does not shape the learned representation.
At inference, the shared head is applied after each of the six rollout steps, producing deterministic forecasts at 7--42-day leads for evaluation in Section~\ref{sec:eval}.

\subsection{Diffusion-based generative model}

Ensemble members are generated by a conditional diffusion model that learns the residual between the verifying truth and the deterministic forecast, with both the backbone and readout head kept frozen.
A separate model is trained for each of the three target variables.

The diffusion model is conditioned on the decoded deterministic forecasts of the three target variables at the same lead, on the latent state from which they were decoded, on the forecast lead, and on the day of year of the target week.
The latent state is the predictor output at that lead: one vector per $6\times6$ patch, formed by concatenating the five level tokens of the patch ($5\times512$ channels), which gives 800 latent tokens on the $20\times40$ patch grid.
The denoising network is a U-Net on the $120\times240$ grid whose two coarsest stages work on the same $20\times40$ grid, so that every feature of these stages corresponds to exactly one latent token.
At these stages each field token attends to all 800 latent tokens through cross-attention.
The diffusion-based generator therefore reads the latent state that produced the forecast, not only the decoded fields.
Following GenCast~\citep{price2025probabilistic}, we adopt the EDM preconditioning and training objective~\citep{karras2022elucidating} with unit data scale.
Each model is trained on 1979--2018, and the checkpoint with the lowest fair CRPS on 2019--2021 is selected.

Ensemble members are sampled using the second-order Heun solver of EDM with 20 steps along a power-law noise schedule.
The resulting ensemble is calibrated separately for each variable and lead on initializations from January 2022 to 6 December 2023, preceding the test period.
Members are centred on the deterministic forecast and their deviations are scaled, in the standardised frame and for precipitation in the logarithmic frame, such that the ensemble spread matches the error of the deterministic forecast.
Calibration is performed in standardized anomaly space and, for precipitation, in logarithmic space, before applying the inverse transformations.
The test ensembles contain 50 members, while the hindcasts used to construct the model's own climatology (Sec.~\ref{sec:eval}) contain 11 members.

\section{Training details}
\label{app:training}

All models were trained on a single NVIDIA H200 GPU with PyTorch 2.7, bfloat16 autocast and the AdamW optimiser ($\beta_1=0.9$, $\beta_2=0.95$, weight decay 0.05 for the forecast model and 0.1 for the diffusion heads).
Gradients are clipped to a global norm of 1.
Learning rates follow a warmup--stable--decay schedule: a linear warmup, a constant plateau and a linear decay to zero over the final iterations given in Table~\ref{tab:train-backbone}.
Snapshots are written every 1,000 iterations and the reported model is the snapshot with the best validation skill, never the last one.

Table~\ref{tab:train-backbone} lists the detailed settings of the two training stages.
Stage 1 trains the encoder, the predictor and the readout head on (one step) with the joint-embedding objective: the predicted latent state is compared with the LayerNorm-standardised, stop-gradient encoding of the target week by an exponential-moving-average (EMA) teacher.
The loss is the L1 distance per token, weighted by the cosine of latitude; each latent dimension is rescaled by the inverse of its running standard deviation so that no dimension dominates.
The teacher momentum increases linearly from 0.999 to 1 over the stage.
Stage 2 continues from the Stage-1 weights at iteration 10,000 with a fresh optimiser and rolls the predictor for all $K=6$ steps.
Every step is trained against the teacher encoding of its own target week with the same objective, the latent state is detached between steps, and the teacher is frozen.
The learning rate of the backbone is 25 times smaller than in Stage 1; the readout head keeps its learning rate.
Stage 2 also switches the precipitation target of the readout head to the weekly total in the logarithmic frame.
The validation set is a fixed subset of 1,048 initialisations from 2019--2021 on which the model is rolled out to 42 days every 500 iterations; the Stage-2 snapshot at iteration 2,000 had the best validation skill at weeks 3--6 and is the model used throughout the paper.
\begin{table}[h]
\centering
\caption{Training settings of the forecast model.
Both stages use the same architecture (91.0M trainable parameters including the readout head), the same data and the same optimiser.}
\label{tab:train-backbone}
\small
\begin{tabularx}{\textwidth}{lXX}
\toprule
 & Stage 1: pre-training & Stage 2: rollout fine-tuning \\
\midrule
Initialisation & random & Stage 1, iteration 10,000 \\
Rollout steps per sample & 1 & 6 (leads 7--42 days) \\
Objective & JEPA (L1 in latent space) & JEPA at every step, detached hand-offs \\
Teacher & EMA, momentum 0.999 $\rightarrow$ 1 & frozen \\
Context / target length & 7 / 7 days & 7 / 7 days \\
Batch size & 32 & 32 \\
Learning rate (backbone / readout head) & $5\times10^{-5}$ / $10^{-4}$ & $2\times10^{-6}$ / $10^{-4}$ \\
Schedule (warmup / total / decay) & 1,000 / 20,000 / 2,000 & 300 / 4,000 / 400 \\
\bottomrule
\end{tabularx}
\end{table}

Table~\ref{tab:train-edm} lists the training settings of the diffusion model, which are shared by the three heads except for the noise distribution, whose mean and spread were chosen per variable on the validation years.
Each head is trained on rollouts of the frozen S2S-JEPA model with the residual normalised per lead and variable, and the snapshot with the lowest fair CRPS on the validation years is kept.
Members are drawn with the Heun sampler of EDM and calibrated as described in Appendix~\ref{app:model}.
\begin{table}[h]
\centering
\caption{Training settings of the diffusion model (one head per variable, 9.3M parameters each).}
\label{tab:train-edm}
\small
\begin{tabularx}{\textwidth}{lX}
\toprule
Setting & Value \\
\midrule
Noise distribution $(P_\mathrm{mean}, P_\mathrm{std})$ & T2M $(-0.6, 1.4)$; MSLP $(0.5, 1.7)$; TP $(-1.8, 1.0)$ \\
Data scale $\sigma_\mathrm{data}$ & 1 (residual normalised per lead and variable) \\
Training years / validation years & 1979--2018 / 2019--2021 \\
Batch size & 32 \\
Learning rate & $2\times10^{-4}$, warmup 200, decay over the last 600 of 3,000 iterations \\
Weight decay & 0.1 \\
EMA of weights & 0.999 \\
Validation & every 250 iterations, 32 initialisations, 8 members, 10 sampling steps \\
Selection criterion & lowest fair CRPS \\
Sampling & Heun, 20 steps, $\sigma$ from 80 to 0.002, $\rho=7$ \\
Members & 50 (test), 11 (hindcasts) \\
\bottomrule
\end{tabularx}
\end{table}

\section{Significance Test}\label{app:significance}

Differences between two systems are assessed with a paired moving-block bootstrap over the 102 initialisation dates.
Weekly initialisations are not independent, because target weeks overlap and consecutive initialisations share the same slowly varying background, so resampling single dates would overstate the effective sample size.
We therefore resample blocks of eight consecutive initialisations with replacement until 102 dates are reached, and use the same resampled dates for both systems, so that the difference is evaluated on identical weather.
For the TCC the per-grid-point correlation is recomputed on each resample before the area-weighted mean is taken; for the RPSS the per-initialisation area-mean RPS of both systems and of the uniform forecast are resampled and the skill scores recomputed, with climatologies, trends and calibration kept fixed.
Each test uses 1,000 resamples; a difference is called significant when the 2.5--97.5 percentile interval of its bootstrap distribution excludes zero.
Blocks of eight weeks absorb sub-seasonal autocorrelation but not interannual variability; with two test years the intervals should be read as optimistic.

\section{Trend removal in the probabilistic evaluation}\label{app:trend}

\begin{figure}[h!]
\centering
\includegraphics[width=0.95\linewidth]{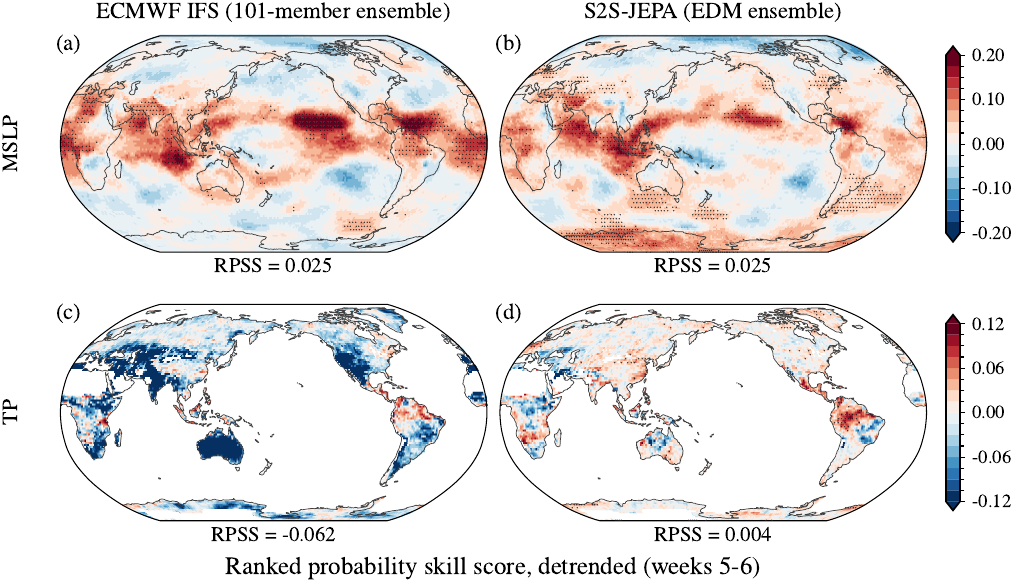}
\caption{As in Fig.~\ref{fig:rpss}, but with system-specific linear trends removed for MSLP and TP.}
\label{fig:rpss_detr}
\end{figure}

Following the protocol of FuXi-S2S, a linear trend can be removed from forecasts and observations before the quintile categories are formed, so that no skill is awarded for reproducing the warming of the reference period.
The main text applies this step to T2M only, because its trend is large and well estimated, whereas the MSLP trend is negligible and the per-grid-point precipitation trend is dominated by sampling noise.
Figure~\ref{fig:rpss_detr} shows the same evaluation with the trend removed for MSLP and TP as well.
Here, the scores shift in favour of S2S-JEPA, mainly because the ECMWF scores decrease markedly; we attribute this to the noise of the extrapolated precipitation trend rather than to forecast skill and therefore do not use it in the main text.

\end{document}